\documentclass[9pt,sigconf,letterpaper]{acmart}
\newcommand{\subhead}[1]{\noindent{\textbf{#1.}}}

\usepackage{balance}
\usepackage{xspace, xparse}

\NewDocumentCommand\mm{g}{%
  \IfNoValueF{#1}{{\color{blue} \textbf{(MM: #1)}}}%
  \IfNoValueT{#1}{{\color{blue} \textbf{(MM)}}}%
}
\definecolor{darkgreen}{RGB}{0,100,0}
\NewDocumentCommand\tr{g}{%
  \IfNoValueF{#1}{{\color{darkgreen} \textbf{(TR: #1)}}}%
  \IfNoValueT{#1}{{\color{darkgreen} \textbf{(TR)}}}%
}

\setcopyright{none}
\renewcommand\footnotetextcopyrightpermission[1]{} 
\acmConference{}
\acmBooktitle{}
\acmYear{2026}
\acmMonth{2}

\acmDOI{}  
\acmISBN{} 

\author{Tarek Ramadan}
\email{tarek.ramadan@concordia.ca}
\affiliation{
  \institution{Concordia University}
  \city{Montreal}
  \state{QC}
  \country{Canada}
}

\author{AbdelRahman Abdou}
\email{abdou@scs.carleton.ca}
\affiliation{
  \institution{Carleton University}
  \city{Ottawa}
  \state{ON}
  \country{Canada}
}

\author{Mohammad Mannan}
\email{m.mannan@concordia.ca}
\affiliation{
  \institution{Concordia University}
  \city{Montreal}
  \state{QC}
  \country{Canada}
}

\author{Amr Youssef}
\email{amr.youssef@concordia.ca}
\affiliation{
  \institution{Concordia University}
  \city{Montreal}
  \state{QC}
  \country{Canada}
}

\usepackage{pgfplots}
\pgfplotsset{compat=1.18}  
\tikzstyle{step} = [rectangle, rounded corners, minimum width=5cm, minimum height=1.2cm, text centered, draw=black, fill=gray!10]
\tikzstyle{arrow} = [thick,->,>=stealth]
\usepackage{tikz}
\usetikzlibrary{shapes.geometric, arrows.meta, positioning}
\usepackage{longtable}
\usepackage{booktabs}

\usepackage{multirow} 
\usepackage{tabularx}
\usepackage{placeins}
\usepackage{tikz}
\usepackage{amsmath}
\usepackage{hyperref}
\usepackage{url}
\usepackage{pgf-pie}
\usepackage{url}
\usepackage[dvipsnames]{xcolor} 
\usepackage{titlesec}
\usepackage{float}
\titlespacing*{\subsubsection}{0pt}{2pt plus 1pt minus 1pt}{2pt plus 1pt minus 1pt}
\usepackage{enumitem}

\usepackage{xparse}
\usepackage{xspace}

\usetikzlibrary{shapes.geometric, arrows.meta, positioning, shapes.multipart}

\usepackage{xcolor}

\definecolor{darkred}{rgb}{0.8, 0, 0} 
\newcommand{\black}[1]{\textcolor{black}{#1}}

\tikzstyle{process} = [rectangle, minimum width=2.5cm, minimum height=0.8cm, text centered, draw=black, fill=blue!10]
\tikzstyle{arrow} = [thick,->,>=Stealth]

\newcommand{\availabilityDiscardRate}{82\%}
\newcommand{\liveURLCount}{149{,}450}
\newcommand{\suspiciousKeywordCountHTML}{402}
\newcommand{\suspiciousKeywordCount}{302}

\newcommand{\perctageforinclauth}{43.3\%}

\newcommand{\perctageforinclfinance}{19.6\%}

\newcommand{\inclusiveAuthCount}{5343}

\newcommand{\conservativeTokenCount}{2442}       
\newcommand{\conservativeVerifiCount}{347}

\newcommand{\inclusivePasswordCount}{1,117}

\newcommand{\inclusiveFinancialCount}{2412}

\newcommand{\inclusivePromoCount}{313}

\begin{document}

\title[Measuring Sensitive Data Leaks Across Public URL Repositories]{The Silent Spill: \\Measuring Sensitive Data Leaks Across Public URL Repositories}



\begin{abstract}
A large number of URLs are made public by various platforms for security analysis, archiving, and paste sharing---such as VirusTotal, URLScan.io, Hybrid Analysis, the Wayback Machine, and RedHunt. These services may unintentionally expose links containing sensitive information, as reported in some news articles and blog posts. However, no large-scale measurement has quantified the extent of such exposures. We present an automated system that detects and analyzes potential sensitive information leaked through publicly accessible URLs. The system combines lexical URL filtering, dynamic rendering, OCR-based extraction, and content classification to identify potential leaks. We apply it to 6,094,475 URLs collected from public scanning platforms, paste sites, and web archives, identifying 12,331 potential exposures across authentication, financial, personal, and document-related domains. These findings show that sensitive information remains exposed, underscoring the importance of automated detection to identify accidental leaks.
\end{abstract}


\maketitle

\section{Introduction}
Over time, URLs have evolved from simple online resource locators into carriers of sensitive data such as authentication tokens, document identifiers, and recovery links. 
Thus, a single leaked URL can expose an entire account, a signed contract, or an organization’s internal data. Across the internet, URLs are constantly shared, scanned and stored by security analysis services, repositories, and paste sites. Some of these links may lead to private files, login tokens, password-reset pages, or personal/organizational documents. Understanding what these URLs contain and how they spread is key to seeing the real risk behind them. 

Several web services, such as VirusTotal~\cite{virustotal}, Hybrid Analysis~\cite{hybrid} and URLScan.io~\cite{urlscan}, scan user-submitted URLs for security issues, e.g., malware and phishing. Email providers and enterprise security gateways also often automatically forward (suspicious) URLs to these security scanners for analysis. These platforms often leave submitted URLs visible in public reports (e.g., for further look-up without requiring a new analysis).  Other sources, such as the Wayback Machine~\cite{wayback} and RedHunt paste sites~\cite{redhunt}, also host and archive URLs that remain online long after use. These public URLs, however, may unintentionally expose private data.
Examples include: 
private links revealing API tokens, and private GitHub pages and internal URLs~\cite{positive2022urlscan}; and cloud/NAS-hosted files, 
corporate communication, and Oauth sign-in links~\cite{vin01-blog}. 

Analyzing leaks from public URLs at scale presents, however, several practical challenges. URLs often expire or redirect, and public datasets from scanning platforms and web archives frequently contain redundant or benign entries. Sensitive content may also appear only after dynamic rendering, and web content exists in diverse formats---HTML, PDFs, images, and dynamically generated elements---requiring multiple extraction methods. Maintaining accuracy across millions of URLs requires balancing strict filters to reduce false positives with flexible analysis to capture potential leaks.

We developed a lightweight detection system that combines lexical analysis, dynamic content inspection, and category-based sensitivity classification. Our goal is to analyze many public sources while maintaining accurate and efficient detection.
We collected 6,094,475 URLs from public scanning services and web archives, then filtered and analyzed them for potential exposures. We found 12,331 potential exposures, including 26 live password-reset links, 83 API keys in query parameters or inline content, and 12 publicly accessible e-signature workflows. We also found persistent 2FA backup codes and unexpired JSON Web Token (JWT) tokens. We notified all repository operators and 53 affected parties, several of whom responded with acknowledgments and fixes.

This paper presents our initial contributions:
\begin{itemize}[leftmargin=2.5em, itemsep=0pt, topsep=0pt]
    \item We build a cross-platform dataset of roughly six million URLs collected from five public sources.
    \item We design and implement a scalable and lightweight methodology\footnote{Source code available at: \url{https://github.com/anonymus400/silent-spill-heuristic-system.git}} for detecting potential URL leaks.
    \item We apply our system, which uses lexical, structural, and dynamic inspection, to measure potential URL-based exposures.
\end{itemize}
\section{Related Work}
\label{sec:background}

Smaragdakis et al.~\cite{smaragdakis2020sensitive} trained a machine-learning classifier on a billion URLs from Common Crawl~\cite{commoncrawl} using Curlie.org\cite{curlie} categories to categorize URLs that belong to five sensitive categories: Ethnicity, Health, Political Beliefs, Religion, and Sexual Orientation. 
Borders and Prakash~\cite{borders2008quantifying} introduced a framework to measure how much information web requests reveal, ignoring predictable details such as browser headers or timestamps that are not actual leaks. 
West and Aviv~\cite{west2018query} developed a tool to measure privacy disclosures in URL query parameters, showing that identifiers and credentials can be exposed through GET requests. They provided a classification of parameter types associated with private data, advancing understanding of URL-based information disclosure.
%
Kim et al.~\cite{kim2022phishing} developed a phishing detection system using structural and lexical features such as token distribution, path depth, subdomain entropy, and parameter naming conventions. Their framework demonstrated how structural and lexical attributes can effectively characterize suspicious URLs.
%
GitHound~\cite{githound} and SecretFinder~\cite{secretfinder} apply regex-based pattern matching to identify secrets in GitHub repositories and JavaScript files. These tools are used in development workflows to detect API keys before release, contributing to automated secret detection practices.

\begin{sloppypar}
\subhead{Research gap} Overall, existing studies examine URL-based data exposure from different perspectives, but do not capture the current scale and nature of URL-based leaks. Smaragdakis et al.~\cite{smaragdakis2020sensitive} performed large-scale classification of sensitive URLs. A sensitive URL and a URL leak are quite different. A URL is sensitive if it falls under any of the five categories above~\cite{smaragdakis2020sensitive}. A URL leak provides direct public access to a private object that should not be publicly accessible. For example: \texttt{example.org/cancer-support/} is a sensitive URL because it is health related, but we do not consider this a leak herein, whereas \texttt{exa.com/invoice.pdf?acs\_tkn=AF3} is a URL leak, but it would not fall under the defined sensitive categories. 
\black{West and Aviv~\cite{west2018query} analyzed query-string disclosures and built a classification of sensitive parameters, but did not cover dynamic or archived sources.}
Borders and Prakash~\cite{borders2008quantifying} introduced an early framework quantifying information flow in HTTP requests, but did not target concrete data leaks.
GitHound~\cite{githound} and SecretFinder~\cite{secretfinder} detect exposed keys in code repositories, but do not cover live or dynamic web analysis.
Despite these advances, the scale, persistence, and cross-platform nature of URL leaks remain unmeasured, motivating a systematic detection and measurement effort across security analysis, archival, and sharing platforms using dynamic rendering and large-scale inspection. We address this gap herein.

\end{sloppypar}

\begin{table}[h]
\centering
\footnotesize
\vspace{-2pt}
\caption{Static-resource types excluded in Stage~1. PDFs are retained for text/OCR.}
\vspace{7pt}

\begin{tabular}{@{}p{1.6cm}p{3.6cm}p{2.3cm}@{}}
\toprule
\textbf{Category} & \textbf{Extensions (examples)} & \textbf{Handling} \\ 
\midrule
Stylesheets & \texttt{.css} & Exclude (static) \\
Fonts & \texttt{.woff}, \texttt{.woff2}, \texttt{.ttf}, \texttt{.otf}, \texttt{.eot} & Exclude (static) \\
Icons & \texttt{.ico} & Exclude (static) \\
Images & \texttt{.png}, \texttt{.jpg}, \texttt{.jpeg}, \texttt{.gif}, \texttt{.bmp}, \texttt{.svg}, \texttt{.webp} & Exclude (static), \emph{except} whitelisted domains\footnote{Image, paste, and e-signature domains are retained for screenshot/OCR as described in \S3.2.} \\
Media & \texttt{.mp3}, \texttt{.mp4}, \texttt{.mov}, \texttt{.avi} & Exclude (static) \\
PDFs & \texttt{.pdf} & \textbf{Retain} (text/OCR) \\
\bottomrule
\end{tabular}

\vspace{-6pt}
\label{tab:static-ext}
\end{table}




\section{Our Framework}
\label{sec:method}

To systematically identify sensitive data exposed via public URLs, we present an automated detection system that efficiently processes large datasets and detects potential leaks embedded in HTML structure, metadata, or URL parameters by combining keyword-based filtering with content analysis across rendered pages, structured fields, and OCR extraction. \autoref{fig:pipeline} summarizes the overall workflow. We detail each stage below.

\begingroup
\subsection{Initial List of URLs}
To build a diverse dataset of potentially sensitive URLs, we collect unique entries between January and April 2025 from five sources covering real-time analysis platforms and historical archives: VirusTotal, Hybrid Analysis, URLScan.io, Wayback Machine, and Redhunt paste sites. URLs are collected from each source as follows.

(1) \textit{URLScan.io~\cite{urlscan}:}
We use domain- and keyword-based queries to retrieve 1,000-URL batches via the Search API,\footnote{\url{https://urlscan.io/docs/api/}} contributing 5,515,585 URLs in total.
(2) \textit{VirusTotal~\cite{virustotal}:}
We query the VirusTotal API\footnote{\url{https://developers.virustotal.com/reference/overview}} across risk-prone domains (\autoref{tab:high-risk-domains}, in the appendix) frequently used for payments, authentication, and document sharing, yielding 132,919 URLs.
(3) \textit{Wayback Machine~\cite{wayback}:}
We use the CDX API\footnote{\url{https://web.archive.org/cdx/search/cdx}} to collect archived URLs, contributing 363,550 entries.
Queries use prefix matching (e.g., \texttt{.example.com/}) with JSON output, retaining accessible URLs across domain groups (\autoref{tab:wayback-domain-sources}, in the appendix).
(4) \textit{Hybrid Analysis~\cite{hybrid}:}
We extract URLs from public sandbox reports via the open feed,\footnote{\url{https://www.hybrid-analysis.com/search?query=}} processing 3,000 reports-which represent all publicly accessible reports during the collection period-yielding 68,892 URLs.
(5) \textit{RedHunt Paste Sites~\cite{redhunt}:}
We scan publicly available paste-sharing domains through live scraping and archived index pages,\footnote{\url{https://redhuntlabs.com/online-ide-search/}} collecting 13,529 URLs (\autoref{tab:paste-domains}, in the appendix).

The initial dataset contains 6,094,475 URLs (see the first three columns of \autoref{tab:source-composition}). 
This dataset serves as the input to our four-stage leak-detection system.

\begin{figure}[t]
    \centering
    
    \includegraphics[width=\columnwidth]{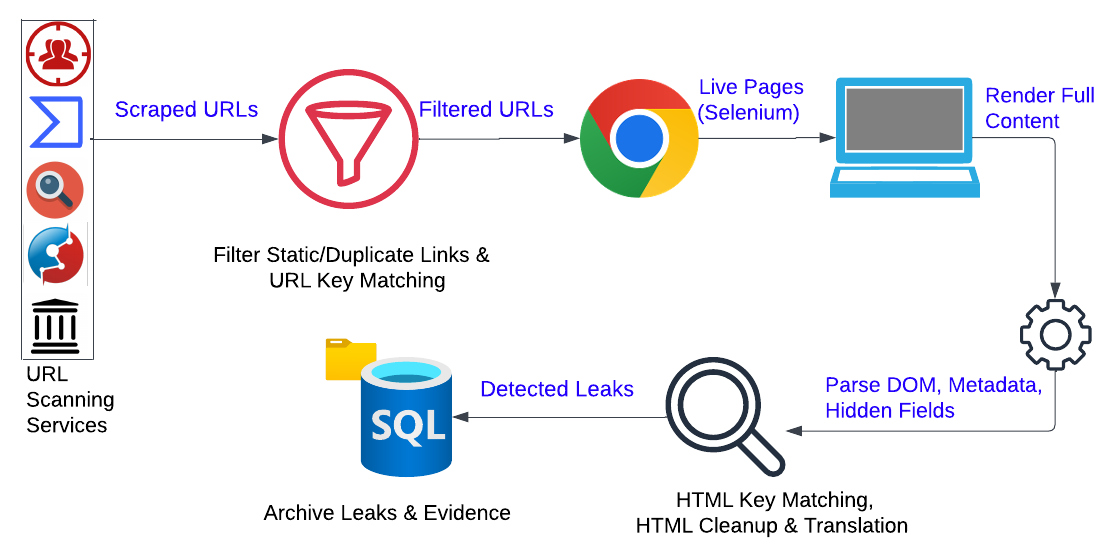}
    \caption{Leak Detection Workflow Overview}
    \label{fig:pipeline}
\end{figure}

\subsection{Stage 1: URL Filtering}
\label{sec:url-filtering}
We begin by filtering non-HTML static resources such as stylesheets, fonts, and icons (e.g., \texttt{.css}, \texttt{.woff}, \texttt{.ico}); however, we keep \texttt{.pdf} files, as they may contain sensitive text or embedded data. Static file types are listed in Table~\ref{tab:static-ext}. We also retain both HTTP and HTTPS links. 
After removing static resources, 4,982,613 URLs remained.
We keep URLs from image-sharing, paste, and e-signature platforms regardless of file type to maintain domain coverage, and we store both screenshot and HTML copies for later analysis.

We curate \suspiciousKeywordCount{} identifiers (Tables~\ref{tab:suspicious-params-part1} and~\ref{tab:suspicious-params-part2}, in the appendix) used for URL-parameter leak detection, aggregated from prior work~\cite{west2018query,smaragdakis2020sensitive}, public datasets, and live web fields. The list includes security terms (e.g., \texttt{auth}, \texttt{token}, \texttt{session}) and operational markers (e.g., \texttt{invoice}, \texttt{booking}, \texttt{tracking}) to detect both security and business-related exposures.
We then perform lexical matching on two parts of each URL: (1) the path segments, and (2) the query parameters and their values.
For example, in:
\url{https://www.example.com/aa/bb?var1=foo\&var2=bar},
the path segments are \texttt{aa},\texttt{ bb}, and the parameters are \texttt{var1= foo}, \texttt{var2=bar}. Each element is checked for substrings that match any of our \suspiciousKeywordCount{} identifiers. Lexical matching reduced the dataset to 2,104,376 URLs.

We also retain URLs meeting any of the three structural criteria:
(i) URLs containing at least five parameters or field names longer than
seven characters; (ii) a total length exceeding 150 characters;
and (iii) four or more directory levels, common in nested or
internal resources. We set these thresholds after testing them
on a 10K-URL pilot dataset, where 7,842 URLs (78.42\%) showing potential leaks followed these patterns. In a manual review of 100 URLs from the retained set and 100 from the remainder, 92 retained URLs contained leaks, compared to 7 in the non-retained group. 
Finally, short or single-parameter URLs (e.g., \texttt{?id=123}, \texttt{?t=abc}) are retained for entropy-based analysis to detect random tokens or identifiers. This ensures concise but potentially sensitive URLs are not discarded. Applying the structural criteria further reduced the set to 832,714 URLs.

Combined, Stage 1 reduces the initial 6,094,475 URLs to 832,714 candidates. Table~\ref{tab:url-filtering-summary} shows the reduction in URL counts after each stage.

\subsection{Stage 2: Retrieve Content}
Each URL passing lexical filtering undergoes live validation to confirm reachability and responsiveness. We send an HTTP HEAD request via Selenium; URLs responding with status 200–399 are marked live and proceed to a full GET request to fetch the page body. URLs that return client or server errors, timeout, or fail to respond are skipped.

To handle redirections safely, we allow up to three consecutive redirects, based on a random test of 10,000 URLs in which 8,976 (89.76\%) of valid pages resolved within three hops. This threshold is chosen to ensure consistent coverage without excessive crawling. This approach captures final destinations (e.g., expanded URL shorteners) that may contain sensitive data.
We exclude live URLs displaying messages such as ``Access Denied,'' ``Authentication Required,'' or ``404 Not Found,'' (see Table~\ref{tab:access-denial-keywords}, in the appendix). This ensures that only content-rich, accessible pages proceed to further analysis. Overall, this availability check eliminates $\sim$\availabilityDiscardRate{} of the filtered candidates, yielding \liveURLCount{} live and content-rich URLs. 

\subsection{Stage 3: Render and Extract}
To capture exposures requiring client-side rendering, each candidate URL is loaded in a headless browser. 
The browser loads dynamically generated content, including JavaScript-injected elements, and auto-filled fields that are not available in static HTML. 
Client-side scripts are allowed to run, and the viewport expands to full scrollable height to ensure all rendered content is visible. This step reveals information that appears within a 5-second render window and is absent from the initial static HTML response. We chose 5 seconds as longer renders (10 seconds) revealed less than 1.96\% additional leaks across a random test of 10,000 URLs.
During rendering, visible non-English text is translated via the Google Translate API~\cite{googletranslate}.
After rendering, we perform a structural inspection to prepare each page for detailed content analysis. 
We analyze the full DOM to capture JavaScript-injected elements, dynamic forms, and HTML input fields marked as \texttt{type=hidden}, which store data not visible on the rendered page. We parse embedded metadata (e.g., \texttt{data-*} attributes and meta tags) to surface tokens or document identifiers that may not appear in visible text. 
We also use Tesseract OCR~\cite{tesseract_ocr} to extract text from images so that information visible only in images is also analyzed. We apply OCR or deeper inspection when a page shows very little readable text (e.g., mostly images or icons), or when the URL suggests hidden sensitive data (e.g., reset or token links). This ensures all observable textual, structural, and visual content is available for subsequent analysis.

\subsection{Stage 4: Content Analysis}

At this stage, we analyze page content to detect potential leaks. 
We sanitize the HTML by removing non-content elements such as \texttt{<script>} and \texttt{<style>}.
We analyze the remaining content across three data types: visible text, input field values, and metadata attributes.
We evaluate all extracted values against \suspiciousKeywordCountHTML{} curated identifiers (Table~\ref{tab:full-suspicious-params}, in the appendix), extending the Stage 1 parameter set with additional context terms derived from observed leaks, public datasets, and metadata patterns.
Keyword-based filtering provides the initial candidate set, and we perform structural and contextual checks to confirm if matches represent potential leaks.

\begin{table}[t]

\caption{Reduction and filtering stages applied to the original dataset}

\centering
\small  
\setlength{\tabcolsep}{2.6pt}  
\begin{tabular}{l|c|c|c}
\toprule
\textbf{Stage} & \textbf{Remaining} & \textbf{Eliminated} & \textbf{\% Eliminated} \\
\hline
Initial dataset           & 6,094,475 & –        & – \\
Filter target URLs       &   832,714 & 5,261,761 & 86.3\% \\
Availability check        &   149,450 &   683,264 & 82.03\% \\
Confirmed leaks            &   12,331  &   137,119 & 91.75\% \\
\bottomrule
\end{tabular}
\vspace{-2pt}
\label{tab:url-filtering-summary}
\end{table}

\section{Results}
\label{sec:results}
We now present the leak distribution by category, and highlight our critical findings (for an overview, see Table~\ref{tab:source-composition} and Figure~\ref{fig:leak-barplot}).

\setlength{\tabcolsep}{4.5pt}
\begin{table}[h]
\caption{Detected Leaks Across URL Sources}
\label{tab:source-composition}

\centering
\footnotesize
\begin{tabular}{l|rr|rr}
\toprule

\textbf{Source} & \textbf{\# URLs} & \textbf{\% of dataset} & \textbf{Leak Matches} & \textbf{\% of Total} \\
\midrule
URLScan.io           & 5,515,585 & 90.48\% & 7,826                & 63.5\% \\
Wayback Machine      &   363,550 & 5.97\% & 1,806      & 14.6\% \\
VirusTotal           &   132,919 & 2.18\% & 1,445   & 11.7\% \\
Hybrid Analysis      &    68,892 & 1.13\% & 963                   & 7.8\% \\
Redhunt paste sites &    13,529 & 0.22\% & 291                   & 2.4\% \\\midrule
\textbf{Total}       & 6,094,475 & 100\% & 12,331 & 100\%\\\bottomrule

\end{tabular}
\vspace{-1em}
\end{table}

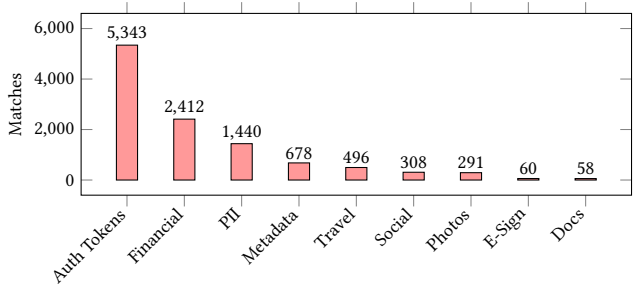
\begin{figure}[h]
\centering
\footnotesize
\caption{Distribution of Flagged Leak Categories.}

\begin{tikzpicture}
\begin{axis}[
    ybar,
    bar width=8pt,
    width=1.05\linewidth,
    height=4cm,
    enlargelimits=0.1,
    ylabel={Matches},
    ylabel style={font=\footnotesize},
    yticklabel style={font=\footnotesize},
    symbolic x coords={
        Auth Tokens,
        Financial,
        PII,
        Metadata,
        Travel,
        Social,
        Photos,
        E-Sign,
        Docs
    },
    xtick=data,
    xticklabel style={rotate=45,anchor=east,font=\footnotesize},
    nodes near coords,
    every node near coord/.append style={
        font=\footnotesize,
        anchor=south,
        yshift=-1pt,
        color=black,
    },
    ymin=0,
    ymax=6000,
]
\addplot[fill=red!40] coordinates {
    (Auth Tokens,5343)
    (Financial,2412)
    (PII,1440)
    (Metadata,678)
    (Travel,496)
    (Social,308)
    (Photos,291)
    (E-Sign,60)
    (Docs,58)
};
\end{axis}
\end{tikzpicture}
\vspace{-10pt}
\vspace{-10pt}
\label{fig:leak-barplot}
\end{figure}

\subsection{Leak Distribution by Category}

Our system processed 6,094,475 URLs end-to-end in 3.8 hours (Ubuntu 24.04.2 LTS on x86\_64, Intel Core i9-10900, Linux kernel 6.14.0, 64 GB RAM, 2 TB HDD), identifying 12,331 potential URL-based leaks, grouped into 9 categories representing the most common leak types observed in our dataset. Figure~\ref{fig:leak-barplot} shows the categories and the number of potential leaks flagged in each category. \emph{Authentication \& Tokens} leaks were the most common (\inclusiveAuthCount{}, \perctageforinclauth{}), indicating that credential-related artifacts appear most frequently in the analyzed URLs.
To verify that the flagged content is true positives (TPs), we chose a random sample from each category for manual inspection. Table~\ref{tab:fprates} shows the sample size we took, the number of false positives (FPs) discovered upon manual inspection, the FP rate, and the Wilson score interval at 95\% confidence~\cite{wilson1927probable}. Sample sizes differ slightly across categories because we chose numbers that kept the inspected URLs roughly similar across categories while keeping the manual review manageable. Considering the upper bound (UB) FP rate of 0.26 (26\%) for the Authentication \& Tokens category, the results show that the flagged potential leaks contain $5343 \times 0.74 = 3,954$ TP URL-based leaks (with 95\% confidence).

The verification process followed consistent labeling rules. One author marked a URL as a true positive if it exposed sensitive data (e.g., API key, token, invoice number, or identifier), and as a false positive if it contained placeholders or benign metadata (e.g., \texttt{<meta name=``author'' content=``Admin''>}). A total of 3 ambiguous cases were reviewed by a second author, of which 2 were confirmed as true positives and 1 as a false positive.

For personally identifiable information (PII), examples included exposed email addresses, full names, or phone numbers in visible text or metadata. Social platform leaks included invitation links, chat tokens, and webhook URLs from services such as Discord or Slack. Metadata-related leaks involved internal reference tags or descriptive fields (e.g., \texttt{created\_by=admin}) that revealed organizational context.
Among all categories, credential-related leaks dominated. Many matched fields such as \texttt{authentication\_token} (\conservativeTokenCount{}) and \texttt{verification\_ token} (\conservativeVerifiCount{}), accounting for 43.3\% of all exposures. Financial artifacts ranked second (\inclusiveFinancialCount{}, \perctageforinclfinance{}), commonly matching parameters like \texttt{checkout\_id} (\inclusivePasswordCount{}) and \texttt{payment\_token} (\inclusivePromoCount{}), often embedded in order confirmations or client-side payment flows lacking access control. Such identifiers could expose order details or enable replayed payment requests. Other categories, including document exposures and location artifacts, appeared less frequently.

Most potential leaks (63.5\%) originate from URLScan.io (see the last two columns of Table~\ref{tab:source-composition}), where public scans often reveal API and session URLs submitted for reputation analysis. The Wayback Machine contributes 14.6\% of potential leaks retrieved through its CDX indexing API, which provides access to archived web snapshots. These URLs typically stem from content that was once public—such as invoices, documents, or dashboards—but remains accessible in archived form even after its intended use was completed. VirusTotal accounts for 11.7\%, mainly exposing password-reset links. Hybrid Analysis contributes 7.8\%, while Pastebin yields 2.4\%, primarily from static credentials, configuration data, or file links shared in public pastes.

\begin{table}[h]
\small
\caption{Sample size (SS), False Positives (FP), FP Rate, and Wilson interval (Lower and Upper Bounds).}
\label{tab:fprates}
\setlength{\tabcolsep}{1.5pt}
\begin{tabular}{p{4cm}|ccrrr}
\toprule
\textbf{Category} & \textbf{SS} & \textbf{\# FPs} & \textbf{FP rate} & \multicolumn{2}{c}{\textbf{Wilson Score}}\\
\cline{5-6}
 & & & & \textbf{LB} & \textbf{UB}\\
\hline
Authentication \& Tokens & 30 & 3 & 0.10 & 0.035 & 0.26\\
Personally Identifiable Info & 31 & 4 & 0.13 & 0.052 & 0.29\\
Financial Information & 33 & 4 & 0.12 & 0.047 & 0.27\\
Document Access & 32 & 4 & 0.13 & 0.053 & 0.29\\
Travel \& Booking & 30 & 4 & 0.13 & 0.051 & 0.29\\
Social Platforms \& Messaging & 34 & 5 & 0.15 & 0.066 & 0.30\\
Miscellaneous Metadata & 30 & 4 & 0.13 & 0.051 & 0.29\\
E-Signature Credentials & 12 & 0 & 0 & 0 & 0.24\\
\bottomrule
\end{tabular}

\end{table}

\subsection{Critical Exposure Findings}
Manual validation reveals several recurring noteworthy exposure types, summarized below.

\subhead{Exposed 2FA backup codes}  
Backup codes are static, pre-generated tokens used to bypass two-factor authentication (2FA) when users lose access to their primary method. We identified 7 rendered pages displaying fully visible codes, typically 8–10 characters long, and 16 showing partial masking (e.g., half-hidden digits). Because backup codes are rarely rotated, several may remain valid over time, creating long-term risks until affected accounts are manually resecured. Fully visible codes could allow account takeover, as long as they remain valid.

\subhead{Publicly accessible password reset links}  
We found 26 password reset links appearing in rendered content or as query parameters. For example, one reset URL appears on a social media recovery page and another on a retail checkout subdomain (\texttt{checkout.shopify.com}), both containing one-time tokens that could allow unauthorized resets if accessed. If exploited before expiration, these reset tokens can allow unauthorized password changes, resulting in immediate account compromise.

\subhead{Exposed persistent JSON web tokens (JWTs)}  
JWTs are commonly used to maintain authenticated sessions and are expected to expire after short periods. We identified 62 URLs containing tokens that either lack expiration controls or remain valid for long durations. Inspection of token claims (\texttt{iat}, \texttt{exp}) shows multi-day expiration durations, and several tokens without an \texttt{exp} field. These URLs appear on domains such as \textit{iris.cigna.com} and \textit{identitydev.freedomconnect.com}, which correspond to authentication or staging environments. Such tokens can enable long-term unauthorized access even after logout.

\subhead{Unprotected e-signature workflows}  
We identified 12 active signing sessions hosted on e-signature platforms that are publicly accessible. Examples include order contracts and job offer agreements. These URLs allow unauthorized users to view or modify contracts. Such exposures pose legal and regulatory concerns by undermining the integrity and accountability of digital signing processes.

\section{Limitations} 
First, our TP/FP estimates rely on random sampling within each leak category, assuming a uniform false-positive distribution. In practice, certain domains contributed disproportionately to FP cases (e.g., repeated login templates). Although category-level estimates remain statistically valid, domain-level skew may bias estimates. Second, while stratified manual inspection provides statistically valid confidence intervals for precision, measuring recall at scale remains difficult because no comprehensive, labeled dataset of confirmed leaks exists. Evaluating system coverage without exposing sensitive data therefore remains an open challenge. Third, source imbalance partly reflects differences in platform accessibility. URLScan.io provides broad, self-service access to historical and live results, while other repositories impose stricter access controls (e.g., approval-based access, account vetting, stricter rate limits, restricted historical search APIs, and limited bulk-export functionality). These constraints made large-scale collection uneven across sources. Although leaks were found across all platforms we sampled, future work may prioritize more balanced collection. 

\section{Mitigation and Broader Impact}
\label{appendix:mitigation}

Our findings point to several practical steps that different stakeholders could adopt to reduce the risk of URL-based data leaks. Repository operators such as URLScan.io, VirusTotal, and archival services can apply automated filtering to redact common secret-bearing parameters (e.g., \texttt{apikey}, \texttt{sessionid}), incorporate entropy-based heuristics to hide high-entropy tokens before publication, and shift toward private-by-default policies with tiered access so that unredacted results are visible only to submitters or vetted researchers. Platforms can additionally default public reports to masked values while preserving parameter names and structural context, limiting direct credential exposure while maintaining analytical transparency. However, determining which parameters are truly sensitive is context-dependent, and automated masking may miss secrets embedded in unconventional or encoded formats. As an additional measure, repositories could
introduce delayed-publication controls, allowing time to revoke unintentionally exposed sensitive links before public indexing. 

Service providers and developers can strengthen application design by minimizing the use of GET parameters for credentials, employing short-lived or single-use tokens for reset and recovery links, masking identifiers in public pages, and enforcing cache-control and strict referrer policies to prevent leakage via browser logs or intermediaries. At the enterprise and end-user level, organizations can configure security tools not to submit sensitive links to public repositories, deploy internal scans to detect secrets in URLs before they leave the corporate network, and monitor public feeds for exposures of their own domains so that leaked links can be revoked promptly. While no single measure can eliminate these risks, adopting layered mitigations across repositories, providers, and users can substantially reduce the prevalence and persistence of URL-based leaks. 

\section{Closing Remarks}
\label{sec:closing}
We showed that web resources directly accessible through their URLs, without additional access controls, are prone to public listing on URL analysis platforms and web archives once indexed. We uncovered thousands of potential private resources accessible through direct URLs that were not intended to be public. This includes 26 live password reset links, 313 payment tokens, and 62 URLs exposing authentication tokens without expiration constraints. We hope that our findings so far highlight the importance of implementing proper access control mechanisms and avoiding blatant reliance on (perceived) obscurity of URLs in protecting sensitive resources.
Although confirmed leaks account for only ~0.2\% of analyzed URLs, their security implications are substantial. At Internet scale, this fraction corresponds to thousands of active reset links, API keys, and contractual documents that could enable account takeover, fraud, or large-scale data exposure. Our results demonstrate that URL leakage is not a marginal phenomenon but a systemic risk: even infrequent cases accumulate into a significant threat landscape requiring proactive mitigation. 

All our analysis was conducted passively on publicly accessible URLs without authentication or state changes, under the safeguards described in \autoref{appendix:ethics}.
To support future development and reproducibility, we release our pipeline code and identifier sets as open source, while withholding raw datasets to avoid exposing sensitive URLs. In line with our ethical safeguards, we conducted responsible disclosure, notifying operators such as URLScan.io, the Internet Archive, and, when appropriate, individual affected organizations. Beyond measurement, our system can be repurposed defensively to help services automatically identify and redact sensitive URLs before publication. Further details on ethical safeguards are provided in \autoref{appendix:ethics}, and mitigation strategies are outlined in Section~\ref{appendix:mitigation}.

\subsection*{Our Efforts are Ongoing} 
While our current methodology successfully uncovered numerous potential leaks, our ongoing work focuses on adaptability, sustainability, and systematic evaluation of efficacy (e.g., precision and recall). 
To better understand temporal behavior, we performed an automated revisit of 2,000 randomly sampled URLs previously flagged as leaks to measure persistence. After 75 days, 38\% (762 URL) remained publicly reachable without authentication. From this set, we manually verified a stratified subset of 200 URLs to confirm that accessible pages continued to expose sensitive content rather than placeholders or error states. 

We are extending the system to better adapt to varying input types and organizational requirements. Different platforms expose data in varied formats: government services may leak structured identifiers through static fields, while commercial sites often embed sensitive data within dynamically rendered components. Detection objectives also differ by use case. For instance, enterprise security teams may prioritize accuracy and low-noise alerts for continuous monitoring, whereas analysts performing large-scale reconnaissance or threat discovery may prefer more flexible parsing and relaxed thresholds to surface weaker or unconventional signals. To support this range, we are refining our parsing logic, keyword sensitivity, and filtering strategies to adapt dynamically to source characteristics and deployment goals.

Leak patterns and exposure formats may also change over time. Web platforms update their structures, parameter names evolve, and new forms of sensitive data emerge. To maintain effectiveness, we are building longitudinal monitoring mechanisms that integrate new exposure examples, update heuristics, and expand pattern sets as needed. These updates may enable the system to keep pace with evolving exposure trends without major architectural changes. In future work, we plan to conduct repeated longitudinal measurements at multiple time intervals to model leak lifespan, understand how quickly leaked URLs become inactive, and determine whether certain platforms or time periods exhibit higher concentrations of exposures. 
To enable scalable and consistent evaluation, we are integrating lightweight machine learning techniques to assist in automated validation of flagged results. These models will not drive detection but will help estimate error rates, identify false-positive patterns, and prioritize likely true positives for review. This approach strengthens our ability to quantify leak prevalence across large datasets without relying solely on sampling or manual inspection. This effort aims to measure detection accuracy more systematically by automating result validation, improving consistency across evaluation batches, and enabling rigorous large-scale analysis. These improvements will provide deeper insight into system performance as it continues to evolve.

\bibliographystyle{ACM-Reference-Format}  
\bibliography{allsources}

\section*{Appendix}

\appendix
\begingroup

\section{Ethics}
\label{appendix:ethics}
{\color{black}

Assessing the extent of URL-based data leaks raises important ethical considerations. We now outline the safeguards and procedures adopted in this study.

\subhead{IRB guidelines}
Prior to conducting this study, we consulted with the Research Ethics Office at our institution to determine whether formal Institutional Review Board (IRB) review was required. We described the project scope, including the collection and analysis of URLs obtained from public URL scanning services and other publicly accessible repositories. We clarified that the study involved no interaction with human subjects, no attempts to bypass access controls, and no access to non-public systems or authenticated resources. 
Based on this description, the Research Ethics Office at our institution determined that formal ethics review was not required, as the study analyzes publicly accessible information and does not involve human subjects research.

\subhead{Publicly accessible data}
All URLs analyzed in this study originated from publicly accessible sources, including VirusTotal, the Wayback Machine, URLScan.io, Hybrid Analysis, and paste sites. All content was accessed passively via public URLs without requiring credentials or login.

\subhead{Passive leak detection and classification}
Our methodology remained strictly observational: we did not activate password-reset flows, use exposed API keys, or interact with any state-changing elements. For document URLs, we only performed passive \texttt{GET} retrieval of files directly exposed by the public URL, without authentication or form submission. No POST requests, credential submissions, or access-control bypasses were used. Leak classification relied solely on structural patterns, page context, token formats, and visible elements.

\subhead{Validating assumptions about side effects}
To minimize the possibility of unintended side effects, we verified with self-controlled accounts that visiting reset or backup-code links did not invalidate credentials or generate new codes. To reflect the real URLs in our dataset, we enumerated the password-reset and 2FA URL patterns observed in our corpus and replicated these patterns using our own accounts before testing their behavior. We tested 24 distinct password-reset and 2FA URL patterns (covering every unique pattern identified across the 49 exposed URLs). Retrieval logs confirmed the absence of POST requests or other state-changing actions.

\subhead{Limited manual inspection}
Manual inspection was restricted to a small stratified subset of URLs for validation. A local password-protected interface displayed only the URL, matched value, screenshot, and HTML snapshot, enabling consistent labeling while keeping sensitive material confined to an isolated workstation. Sensitive artifacts were anonymized in all reporting.

\subhead{Data storage}
All collected data was stored on an encrypted, passw-ord-protected local disk with no cloud backups, physically accessible only to the first author. All data will be securely destroyed at the end of the project.

\subhead{Responsible disclosure}
We contacted all repository operators involved in our dataset; URLScan.io acknowledged receipt and requested the specific leak patterns (e.g., \texttt{session\_id=}, \texttt{auth\_key=}) to deploy fixes on their end. We also contacted 23 individual organizations responsible for identifiable exposures, of which 11 responded. Additionally, we reached out to 30 affected individuals, and 8 confirmed awareness of their reported exposures. No sensitive URLs, tokens, or user-linked data was disclosed publicly, and no external parties were granted access to the dataset.

\subhead{Risk–benefit tradeoff}
The incremental risk introduced by this study was minimal, in our assessment, since all analyzed URLs were already publicly exposed at the time of collection. The potential benefit is substantial: by documenting these risks and detailing appropriate safeguards, our work supports mitigation efforts by service operators and offers a reproducible framework for future research.

\section{Supplementary Data}\balance
~\autoref{tab:suspicious-params-part1} and~\autoref{tab:suspicious-params-part2} list the lexical identifiers used for URL parameter filtering. ~\autoref{tab:full-suspicious-params} contains 402 structural and semantic keywords derived from file paths, filenames, and artifacts (e.g., invoice, contract, boarding pass). 

~\autoref{tab:suspicious-params-part1} and~\autoref{tab:suspicious-params-part2} together contain 302 lexical identifiers aggregated from prior work~\cite{west2018query,smaragdakis2020sensitive}, public datasets, and common sensitive fields (e.g., \texttt{auth\_token}, \texttt{sessionid}, \texttt{apikey}). 

The two sets partially overlap (e.g., invoice), but one targets structural artifacts while the other targets parameter names. We include both for completeness.

\onecolumn
\addcontentsline{toc}{section}{Appendix: Domain Categories}

\begin{table}[h]
\centering
\caption{Potential High-Risk Domains Queried via VirusTotal}
\label{tab:high-risk-domains}
\begin{tabular}{l}
\hline
\textbf{Payment Platforms} \\
\hline
paypal.com, stripe.com, squareup.com, checkout.stripe.com, venmo.com, pay.google.com, \\
appleid.apple.com, secure.adyen.com, braintreepayments.com, payoneer.com, payu.com, \\
skrill.com, revolut.com, cash.app, zellepay.com \\
\hline
\textbf{Social Platforms} \\
\hline
facebook.com, messenger.com, instagram.com, whatsapp.com, linkedin.com, twitter.com, x.com, \\
t.me, discord.com, snapchat.com, pinterest.com, reddit.com, tiktok.com, wechat.com \\
\hline
\textbf{Cloud Storage \& File Sharing} \\
\hline
drive.google.com, docs.google.com, onedrive.live.com, dropbox.com, box.com, mega.nz, \\
nextcloud.com, owncloud.com, icloud.com, sharepoint.com, wetransfer.com, transfer.sh, \\
send.firefox.com \\
\hline
\textbf{Developer Services \& Code Repositories} \\
\hline
github.com, gitlab.com, bitbucket.org, npmjs.com, pypi.org, hub.docker.com, maven.org, \\
api.github.com, git.io, cdn.jsdelivr.net, cdnjs.cloudflare.com \\
\hline
\textbf{Authentication Providers} \\
\hline
accounts.google.com, appleid.apple.com, auth0.com, okta.com, onelogin.com, \\
sso.squarespace.com, signin.aws.amazon.com, keycloak.org \\
\hline
\textbf{Travel \& Ticketing} \\
\hline
booking.com, expedia.com, airbnb.com, tripadvisor.com, skyscanner.com, hotels.com, \\
trainline.com, ticketmaster.com, eventbrite.com, stubhub.com, checkmytrip.com, \\
ryanair.com, boardingpass.ryanair.com \\
\hline
\textbf{E-Commerce Platforms} \\
\hline
amazon.com, alibaba.com, aliexpress.com, ebay.com, shopify.com, bigcommerce.com, wix.com, \\
woocommerce.com, etsy.com, bestbuy.com, walmart.com, flipkart.com \\
\hline
\end{tabular}
\end{table}
 
\begin{table}[!htbp]
\caption{Wayback Machine Leak Source Domains}
\centering
\label{tab:wayback-domain-sources}
\begin{tabular}{p{0.3\textwidth} p{0.65\textwidth}}
\hline
\textbf{Category} & \textbf{Domains} \\
\hline
Document Sharing & docs.google.com, drive.google.com, app.box.com, mega.nz, pcloud.com, nextcloud.com, icloud.com, filestackapi.com, mediafire.com, transferxl.com, smash.gg, zippyshare.com, anonfiles.com, bayfiles.com, dropbox.com, send-anywhere.com \\
\hline
E-Signature Services & docusign.com, docusign.net, hellosign.com, dropboxsign.com, adobesign.com, secure.adobesign.com, sandbox.esignlive.com, pandadoc.com, signnow.com, eversign.com, signrequest.com, signeasy.com, onespan.com, esignlive.com, zohosign.com, assuresign.com, sertifi.com, legalesign.com, signable.com, formstack.com, signx.wondershare.com \\
\hline
Invoicing / Payment & stripe.com, paypal.com, squareup.com, quickbooks.intuit.com, waveapps.com, zoho.com, bill.com, xero.com, invoiceninja.com, freshbooks.com, invoicehome.com, harvestapp.com, qbo.intuit.com, my.freshbooks.com, zohoinvoice.com \\
\hline
Authentication & auth0.com, accounts.google.com, accounts.google.com/o/oauth2, login.microsoftonline.com, login.live.com, okta.com, \\
& auth.aws.amazon.com, passport.yandex.com, id.heroku.com, bitbucket.org, bitbucket.org/site/oauth2, gitlab.com, gitlab.com/users/sign\_in, \\
& slack.com/signin, firebaseapp.com, appleid.apple.com, console.aws.amazon.com, console.cloud.google.com, \\
& login.salesforce.com, id.atlassian.com, login.cisco.com, auth.digitalocean.com \\

\hline
Paste Sites & pastebin.com, justpaste.it, dpaste.org, hastebin.com, controlc.com, ghostbin.com, paste.ee, rentry.co, zerobin.net, privatebin.net, pastie.io, paste.mozilla.org, termbin.com, snipplr.com, snipt.net \\
\hline
Media Platforms & imgur.com, flickr.com, canva.com, notion.so, figma.com, miro.com, evernote.com, youtube.com/redirect?, slideshare.net, photos.google.com \\
\hline
Developer / Cloud Services & github.com, gitlab.com, bitbucket.org, replit.com, glitch.me, vercel.app, netlify.app, herokuapp.com, render.com, surge.sh, firebaseio.com \\
\hline
Viewers / Converters & docdroid.net, pdfhost.io, scribd.com, docplayer.net, edocr.com, slideserve.com, prezi.com, slidebean.com, docsend.com, view.officeapps.live.com \\
\hline
\end{tabular}
\end{table}


\begin{table}[h]
\centering
\caption{Paste and Code-Sharing Platforms (Redhunt Domains)}
\label{tab:paste-domains}
\begin{tabular}{p{0.9\linewidth}}
\hline
\textbf{General Pastebins and Code Sharing Tools} \\
\hline
paste2.org, pastebin.com, slexy.org, dpaste.org, jsfiddle.net, dotnetfiddle.net, repl.it, \\
paste.pound-python.org, phpfiddle.org, jsbin.com, snipplr.com, hastebin.com, textsnip.com, \\
dpaste.com, ideone.com, paste.opensuse.org, paste.lisp.org, ide.geeksforgeeks.org, \\
play.golang.org, snipt.net, dumpz.org, bitpaste.app, codepen.io, paste.debian.net, \\
rextester.com, paste.xinu.at, pastehtml.com, heypasteit.com, pastebin.fr, codepad.org, \\
justpaste.it, dartpad.dartlang.org, paste.fedoraproject.org, paste.org.ru, \\
try.ceylon-lang.org, paste.frubar.net, paste.ubuntu.com, paste.org, jsitor.com, \\
ide.codingblocks.com \\
\hline
\end{tabular}
\end{table}

\begin{table}[htbp]
\caption{Keywords Used for Access Denial Detection}
\centering
\normalsize
\label{tab:access-denial-keywords}
\begin{tabular}{p{0.85\linewidth}}
\toprule
\textbf{Keyword or Phrase} \\
\midrule
access denied ,
you need permission ,
403 forbidden ,\\
unavailable ,
you have been blocked ,
error 404 ,\\
page doesn't exist ,
404 ,
unauthorized ,
not authorized ,
forbidden ,\\
restricted access ,
private page ,
invalid credentials ,
authentication required \\
\bottomrule
\end{tabular}
\end{table}

\begin{table}
\caption{List of HTML Parameters for Leak Detection}
\centering
\label{tab:full-suspicious-params}
\resizebox{\textwidth}{!}{%
\begin{tabular}{p{0.95\textwidth}}
\hline
\textbf{Authentication \& Session Management} \\
\hline

envelope\_id, signing\_token, sessionid, session\_id, session\_key, sign\_token, doc\_hash, signature\_id, access\_code, signing\_session, approve\_token, verify\_token, auth\_key, esign\_request, contract\_id, legal\_doc, agreement\_id, access\_token, refresh\_token, auth\_token, csrf, xsrf, session\_token, apikey, api\_key, api\_secret, secret\_key, client\_secret, crypto\_key, crypto\_address, private\_key, public\_key, -hash, hmac, security\_code, verification\_code, recovery\_token, reset\_token, oauth\_token, ouath\_verifier, crypto\_wallet, jwt, bearer, state\_token, token, security-login, client\_token, sso\_token, authenticity\_token, csrftoken, logincsrfparam, requestverificationtoken, security\_token \\

\hline
\textbf{User Identity, Social Media \& Personal Identifiers} \\
\hline
national\_id, social\_security\_number, driver\_license\_number, tax\_id, pan\_number, pan\_card, identity\_number, ssn\_last4, health\_id, voter\_id, employee\_id, citizen\_id, residence\_permit, ssn\_full, student\_id, military\_id, beneficiary\_id, email\_address, phone\_number, mobile\_number, fax\_number, emergency\_contact \\

\hline
\textbf{Payment Information \& Transaction Identifiers} \\
\hline
cart\_id, order\_number, invoice\_number, purchase\_id, transaction id, transaction\_token, checkout\_token, order\_reference, payment\_ref, payment\_token, payment\_id, payment\_method, debitcard, creditcard, cvv, bank\_account, stripe\_token, card\_number, transaction\_id, invoice\_id, receipt\_number, account\_number, license\_key, Invoice Number, Invoice ID, Invoice \#, Invoice Total, Invoice Amount, Order Number, Order ID, Order \#, Order Total, Order Amount, Payment Method, Payment Type, Payment ID, Payment Reference, Card Number, Credit Card Number, Card Ending, Cardholder Name, Card Type, Last 4 Digits \\

\hline
\textbf{File Links, Document Access \& Repository Tokens} \\
\hline
fileid, attachment\_id, docid, doc\_token, document\_id, download\_token, gdrive\_file\_id, gdrive\_share\_link, dropbox\_link, onedrive\_link, s3\_bucket, repository\_url, pdf\_file, docx\_file, xlsx\_file, gdrive, onedrive/, Download Link, Document ID, GDrive Link, Dropbox Link \\

\hline
\textbf{Travel, Booking \& Ticketing References} \\
\hline
ticket\_id, reservation\_number, booking\_reference, confirmation\_number, flight\_number, itinerary\_id, passport\_number, trip\_id, e\_ticket\_number, booking\_ref, Boarding Pass, Booking Reference, Reservation Number, Flight Number, Confirmation Number, E-Ticket Number, Itinerary ID, Passport Number, Trip ID \\

\hline
\textbf{Social \& Messaging Integrations} \\
\hline
invite\_link, discord\_invite, discord\_bot\_token, slack\_webhook, slack\_bot\_token, telegram\_bot\_token, telegram\_chat\_id, social\_user\_id, whatsapp\_link, facebook\_id, messenger\_thread\_id, chat.whatsapp, instagram\_id, snapchat\_id, skype\_id, Discord Invite, Slack Invite, Telegram Bot, Messenger Thread, WhatsApp Link, Group ID, Profile ID \\

\hline
\textbf{Address, Contact Details \& Personal Information} \\
\hline
billing\_address, shipping\_address, mailing\_address, home\_address, postal\_code, full\_name, first\_name, last\_name, middle\_name, date\_of\_birth, place\_of\_birth, nationality, marital\_status \\

\hline
\textbf{Medical, Insurance \& Health Records} \\
\hline
insurance\_number, medical\_record\_id, health\_insurance\_id, vaccination\_id, blood\_type \\

\hline
\textbf{Shipping, Tracking \& Logistics Identifiers} \\
\hline
tracking\_id, tracking\_number, survey\_id, event\_id, shipment\_id, parcel\_id, order\_tracking\_id, delivery\_note\_number, waybill\_number, dispatch\_id \\

\hline
\textbf{Corporate, Internal \& Vendor References} \\
\hline
employee\_number, staff\_id, internal\_reference, project\_id, vendor\_id, supplier\_id, purchase\_order\_id, contract\_number, rfq\_id, invoice\_reference \\

\hline
\textbf{Cryptocurrency Wallets \& Blockchain Credentials} \\
\hline
Wallet Address, Crypto Token, Transaction Hash, Mnemonic, Seed Phrase, crypto\_key, crypto\_address, crypto\_wallet \\

\hline
\end{tabular}}
\end{table}

\begin{table}[h]
\centering
\caption{URL Parameters Used for Leak Detection (Part I)}
\label{tab:suspicious-params-part1}
\setlength{\tabcolsep}{2pt}
\renewcommand{\arraystretch}{1.05}
\begin{tabular}{p{0.25\textwidth}p{0.70\textwidth}}
\hline
\textbf{Category} & \textbf{Parameters} \\
\hline

  Authentication \& Session Management & 
 
envelope\_id,\allowbreak  signing\_token,\allowbreak  sessionid,\allowbreak  session\_id,\allowbreak  session\_key,\allowbreak  sign\_token,\allowbreak  doc\_hash,\allowbreak  signature\_id,\allowbreak  access\_code,\allowbreak  signing\_session,\allowbreak  approve\_token,\allowbreak  verify\_token,\allowbreak  auth\_key,\allowbreak  esign\_request,\allowbreak  contract\_id,\allowbreak  legal\_doc,\allowbreak  agreement\_id,\allowbreak  access\_token,\allowbreak  refresh\_token,\allowbreak  auth\_token,\allowbreak  csrf,\allowbreak  xsrf,\allowbreak  session\_token,\allowbreak  apikey,\allowbreak  api\_key,\allowbreak  api\_secret,\allowbreak  secret\_key,\allowbreak  client\_secret,\allowbreak  verification,\allowbreak  verify,\allowbreak  private\_key,\allowbreak  public\_key,\allowbreak  hash,\allowbreak  hmac,\allowbreak  security\_code,\allowbreak  verification\_code,\allowbreak  recovery\_token,\allowbreak  reset\_token,\allowbreak  oauth\_token,\allowbreak  oauth\_verifier,\allowbreak  jwt,\allowbreak  bearer,\allowbreak  state\_token,\allowbreak  token,\allowbreak  security\_login,\allowbreak  client\_token,\allowbreak  sid,\allowbreak  ssid,\allowbreak  sso\_token,\allowbreak  authenticity\_token,\allowbreak  csrftoken,\allowbreak  logincsrfparam,\allowbreak  requestverificationtoken,\allowbreak  security\_token \\ \hline

  API Keys, Webhooks \& Developer Credentials & 
 
webhook\_url,\allowbreak  webhook\_token,\allowbreak  slack\_webhook,\allowbreak  slack\_token,\allowbreak  slack\_bot\_token,\allowbreak  discord\_invite,\allowbreak  discord\_bot\_token,\allowbreak  telegram\_bot\_token,\allowbreak  telegram\_chat\_id,\allowbreak  github\_token,\allowbreak  gitlab\_token,\allowbreak  jenkins\_token,\allowbreak  circleci\_token,\allowbreak  ci\_build\_token,\allowbreak  build\_hook,\allowbreak  api\_webhook,\allowbreak  firebase\_token,\allowbreak  twilio\_sid,\allowbreak  twilio\_token,\allowbreak  zendesk\_token,\allowbreak  zoom\_jwt,\allowbreak  notion\_token,\allowbreak  airtable\_api\_key,\allowbreak  api\_url \\ \hline

  Payment Information \& Transaction Identifiers & 
 
cart\_id,\allowbreak  checkout,\allowbreak  order\_number,\allowbreak  invoice\_number,\allowbreak  purchase\_id,\allowbreak  transaction\_id,\allowbreak  transaction\_token,\allowbreak  checkout\_token,\allowbreak  order\_reference,\allowbreak  payment\_ref,\allowbreak  payment\_token,\allowbreak  payment\_id,\allowbreak  payment\_method,\allowbreak  stripe\_token,\allowbreak  paypal\_token,\allowbreak  square\_token,\allowbreak  debitcard,\allowbreak  creditcard,\allowbreak  cvv,\allowbreak  cvc,\allowbreak  bank\_account,\allowbreak  iban,\allowbreak  bic,\allowbreak  card\_number,\allowbreak  account\_number,\allowbreak  invoice\_id,\allowbreak  receipt\_number,\allowbreak  license\_key,\allowbreak  discount\_code,\allowbreak  promo\_code,\allowbreak  coupon\_code,\allowbreak  voucher\_code,\allowbreak  voucher\_id,\allowbreak  giftcard\_number,\allowbreak  loyalty\_card\_id,\allowbreak  reward\_points,\allowbreak  balance\_amount,\allowbreak  billing \\ \hline

  File Links, Document Access \& Repository Tokens & 
 
file,\allowbreak  file\_url,\allowbreak  fileid,\allowbreak  attachment,\allowbreak  attachment\_id,\allowbreak  doc,\allowbreak  docid,\allowbreak  doc\_token,\allowbreak  document,\allowbreak  document\_id,\allowbreak  download\_id,\allowbreak  download\_token,\allowbreak  gdrive\_file\_id,\allowbreak  gdrive\_share\_link,\allowbreak  dropbox\_link,\allowbreak  onedrive\_link,\allowbreak  s3\_bucket,\allowbreak  repository\_url,\allowbreak  pdf\_file,\allowbreak  docx\_file,\allowbreak  xlsx\_file,\allowbreak  gdrive,\allowbreak  onedrive,\allowbreak  backup\_file,\allowbreak  database\_dump,\allowbreak  config\_file,\allowbreak  logfile,\allowbreak  snapshot\_id,\allowbreak  backup,\allowbreak  config \\ \hline

  Travel, Booking \& Ticketing References & 
 
boardingpass,\allowbreak  visa,\allowbreak  pasabordo,\allowbreak  ticket,\allowbreak  ticket\_id,\allowbreak  reservation\_number,\allowbreak  booking\_reference,\allowbreak  confirmation\_number,\allowbreak  flight\_number,\allowbreak  itinerary\_id,\allowbreak  passport\_number,\allowbreak  trip\_id,\allowbreak  e\_ticket\_number,\allowbreak  booking\_ref,\allowbreak  travel\_doc\_id,\allowbreak  miles\_account\_id,\allowbreak  frequent\_flyer\_number \\ \hline

  User Identity, Social Media \& Personal Identifiers & 
 
invite\_link,\allowbreak  whatsapp\_link,\allowbreak  messenger\_thread\_id,\allowbreak  facebook\_id,\allowbreak  instagram\_id,\allowbreak  snapchat\_id,\allowbreak  skype\_id,\allowbreak  social\_user\_id,\allowbreak  group\_id,\allowbreak  chat\_token,\allowbreak  customer\_id,\allowbreak  client\_id,\allowbreak  user\_id,\allowbreak  account\_id,\allowbreak  member\_id,\allowbreak  profile\_id,\allowbreak  aadhaar\_number,\allowbreak  nhs\_number,\allowbreak  citizen\_id,\allowbreak  residence\_permit,\allowbreak  ssn\_full,\allowbreak  student\_id,\allowbreak  military\_id,\allowbreak  beneficiary\_id,\allowbreak  national\_id,\allowbreak  social\_security\_number,\allowbreak  ssn,\allowbreak  driver\_license\_number,\allowbreak  tax\_id,\allowbreak  pan\_number,\allowbreak  pan\_card,\allowbreak  identity\_number,\allowbreak  ssn\_last4,\allowbreak  health\_id,\allowbreak  voter\_id,\allowbreak  employee\_id,\allowbreak  member\_number,\allowbreak  email\_address,\allowbreak  phone\_number,\allowbreak  mobile\_number,\allowbreak  fax\_number,\allowbreak  emergency\_contact \\ \hline

  Address, Contact Details \& Personal Information & 
 
billing\_address,\allowbreak  shipping\_address,\allowbreak  mailing\_address,\allowbreak  home\_address,\allowbreak  postal\_code,\allowbreak  zip\_code,\allowbreak  full\_name,\allowbreak  first\_name,\allowbreak  last\_name,\allowbreak  middle\_name,\allowbreak  dob,\allowbreak  date\_of\_birth,\allowbreak  place\_of\_birth,\allowbreak  gender,\allowbreak  nationality,\allowbreak  marital\_status,\allowbreak  address \\ \hline

  Medical, Insurance \& Health Records & 
 
insurance\_number,\allowbreak  medical\_record\_id,\allowbreak  health\_insurance\_id,\allowbreak  vaccination\_id,\allowbreak  blood\_type,\allowbreak  healthcare\_id,\allowbreak  dental\_record\_id,\allowbreak  employer\_insurance\_id \\ \hline

  Shipping, Tracking \& Logistics Identifiers & 
 
tracking,\allowbreak  tracking\_id,\allowbreak  tracking\_number,\allowbreak  shipment\_id,\allowbreak  parcel\_id,\allowbreak  order\_tracking\_id,\allowbreak  delivery\_note\_number,\allowbreak  waybill\_number,\allowbreak  dispatch\_id \\ \hline

  Corporate, Internal \& Vendor References & 
 
employee\_number,\allowbreak  staff\_id,\allowbreak  internal\_reference,\allowbreak  project\_id,\allowbreak  vendor\_id,\allowbreak  supplier\_id,\allowbreak  purchase\_order\_id,\allowbreak  contract\_number,\allowbreak  rfq\_id,\allowbreak  invoice\_reference \\ \hline

  Cryptocurrency Wallets \& Blockchain Credentials& 
 
wallet,\allowbreak  wallet\_address,\allowbreak  crypto\_token,\allowbreak  transaction\_hash,\allowbreak  eth\_address,\allowbreak  btc\_address,\allowbreak  mnemonic\_phrase,\allowbreak  seed\_phrase,\allowbreak  keystore,\allowbreak  crypto\_api\_key,\allowbreak  crypto\_secret\_key \\ \hline

\end{tabular}
\end{table}

\begin{table}[h]
\centering
\caption{URL Parameters Used for Leak Detection (Part II)}
\label{tab:suspicious-params-part2}
\setlength{\tabcolsep}{2pt}
\renewcommand{\arraystretch}{1.05}
\begin{tabular}{p{0.25\textwidth}p{0.70\textwidth}}
\hline
\textbf{Category} & \textbf{Parameters} \\
\hline

  Security, Risk \& Audit Identifiers & audit\_log\_id,\allowbreak  incident\_id,\allowbreak  security\_alert\_id,\allowbreak  fraud\_case\_id,\allowbreak  risk\_score,\allowbreak  compliance\_report\_id \\ \hline

  Generic, Broad \& Common Sensitive Terms & 
 
file\_name,\allowbreak  filename,\allowbreak  filepath,\allowbreak  doc\_link,\allowbreak  dataset,\allowbreak  shared\_link,\allowbreak  public\_link,\allowbreak  direct\_link,\allowbreak  temp\_link,\allowbreak  auth,\allowbreak  signin,\allowbreak  login,\allowbreak  creds,\allowbreak  credentials,\allowbreak  password,\allowbreak  pwd,\allowbreak  passwd,\allowbreak  identity,\allowbreak  key,\allowbreak  certificate,\allowbreak  ssh\_key,\allowbreak  oauth,\allowbreak  profile,\allowbreak  account,\allowbreak  settings,\allowbreak  media,\allowbreak  media\_id,\allowbreak  photo,\allowbreak  image,\allowbreak  img\_url,\allowbreak  video\_url,\allowbreak  link \\ \hline

E-Signature Service Domains & esignlive.com,\allowbreak  sandbox.esignlive.com,\allowbreak  docusign.net,\allowbreak  docusign.com,\allowbreak  secure.adobesign.com,\allowbreak  adobesign.com,\allowbreak  hellosign.com,\allowbreak  onespan.com,\allowbreak  signnow.com,\allowbreak  pandadoc.com,\allowbreak  dropboxsign.com,\allowbreak  rightsignature.com,\allowbreak  zohosign.com,\allowbreak  signrequest.com,\allowbreak  eversign.com,\allowbreak  assuresign.com,\allowbreak  formstack.com,\allowbreak  signeasy.com,\allowbreak  sertifi.com,\allowbreak  signable.com,\allowbreak  legalesign.com,\allowbreak  esignly.com,\allowbreak  signx.wondershare.com,\allowbreak  docsketch.com,\allowbreak  getaccept.com,\allowbreak  signaturit.com \\ \hline
 
Hosting Domains & photos.google.com,\allowbreak  lh3.googleusercontent.com,\allowbreak  drive.google.com,\allowbreak  dropboxusercontent.com,\allowbreak  imgur.com,\allowbreak  i.imgur.com,\allowbreak  i.redd.it,\allowbreak  preview.redd.it,\allowbreak  cdn.discordapp.com,\allowbreak  fbcdn.net,\allowbreak  scontent.xx.fbcdn.net,\allowbreak  telegra.ph,\allowbreak  cdn4.telegram-cdn.org,\allowbreak  mmg.whatsapp.net,\allowbreak  onedrive.live.com,\allowbreak  1drv.ms,\allowbreak  s3.amazonaws.com,\allowbreak  bucket.s3.amazonaws.com,\allowbreak  flickr.com,\allowbreak  staticflickr.com,\allowbreak  wetransfer.com,\allowbreak  transfer.sh,\allowbreak  user-images.githubusercontent.com,\allowbreak  prnt.sc,\allowbreak  snag.gy,\allowbreak  gyazo.com,\allowbreak  mail-attachment.googleusercontent.com,\allowbreak  attachments.office.net,\allowbreak  tinypic.com,\allowbreak  imageshack.us,\allowbreak  postimg.cc,\allowbreak  ibb.co,\allowbreak  freeimage.host,\allowbreak  imagevenue.com,\allowbreak  pixhost.to \\ \hline
 
Paste / Code Sharing Domains & paste2.org,\allowbreak jsbin.com,\allowbreak  play.golang.org,\allowbreak  paste.debian.net,\allowbreak  pastehtml.com,\allowbreak  pastebin.com,\allowbreak  snipplr.com,\allowbreak  snipt.net,\allowbreak  heypasteit.com,\allowbreak  pastebin.fr,\allowbreak  slexy.org,\allowbreak  hastebin.com,\allowbreak  dumpz.org,\allowbreak  codepad.org,\allowbreak  jsitor.com,\allowbreak  dpaste.org,\allowbreak  textsnip.com,\allowbreak  bitpaste.app,\allowbreak  justpaste.it,\allowbreak  jsfiddle.net,\allowbreak  dpaste.com,\allowbreak  codepen.io,\allowbreak  dartpad.dartlang.org,\allowbreak  ide.codingblocks.com,\allowbreak  dotnetfiddle.net,\allowbreak  ideone.com,\allowbreak  paste.fedoraproject.org,\allowbreak  paste.frubar.net,\allowbreak  repl.it,\allowbreak  paste.opensuse.org,\allowbreak  rextester.com,\allowbreak  paste.org.ru,\allowbreak  paste.ubuntu.com,\allowbreak  paste.pound-python.org,\allowbreak  paste.lisp.org,\allowbreak  paste.xinu.at,\allowbreak  try.ceylon-lang.org,\allowbreak  paste.org,\allowbreak  phpfiddle.org,\allowbreak  ide.geeksforgeeks.org \\ 
\hline
\end{tabular}
\end{table}

\twocolumn

\end{document}